  \documentclass[multi]{cambridge6Atight}
  \usepackage{natbib}

  \usepackage{rotating}
  \usepackage{floatpag}
  \rotfloatpagestyle{empty}

  \usepackage{amsthm}
  \usepackage{graphicx}
  \usepackage{mathptmx}

  \usepackage{makeidx}
  \makeindex

\copyrightline{This material has been published in \textit{The Impact of Binaries on Stellar Evolution}, Beccari G. \& Boffin H.M.J. (Eds.).
  This version is free to view and download for personal use only. Not for re-distribution, re-sale or use in derivative works. \copyright\ 2018 Cambridge University Press.}

\def\gtsima{$\; \buildrel > \over \sim \;$}
\def\ltsima{$\; \buildrel < \over \sim \;$}
\def\gtrsim{\lower.5ex\hbox{\gtsima}}
\def\lesssim{\lower.5ex\hbox{\ltsima}}

\def\na{\emph{New Astronomy}}
\def\apj{\emph{Astrophysical Journal}}
\def\apjl{\apj{} \emph{Letter}}
\def\nat{\emph{Nature}}
\def\mnras{\emph{MNRAS}}
\def\prd{\emph{Physical Review Documents}}
\def\aap{\emph{Astronomy \& Astrophysics}}

\def\araa{\emph{Annual Review of Astronomy and Astrophysics}}
\def\aj{\emph{Astronomical Journal}}

\begin{document}

%%%%%%%%%%%%%%%%%%%%%%%%%%
%%%% START TO EDIT FROM HERE %%%%%%
%%%%%%%%%%%%%%%%%%%%%%%%%%

  \alphafootnotes
   \author[Michela Mapelli]
          {Michela Mapelli\footnotemark{}$^{,}$\footnotemark{}$^{,}$\footnotemark{}}
  %%%% PUT HERE TITLE OF YOUR CHAPTER  
  \chapter{The Maxwell's demon of star clusters}
  %%% Put A. Einstein et al. in [] if more than 3 authors - these are running authors
  %%% \chapter[Einstein et al.]{Writing a chapter for the Imbase17 book}

%%% footnotes are not compulsory... please use only if needed. Don't put your affiliation here.
  \footnotetext[1]{Institut f\"ur Astro- und Teilchenphysik, Universit\"at Innsbruck, Technikerstrasse 25/8, A--6020, Innsbruck, Austria}
  \footnotetext[2]{INAF -- Osservatorio Astronomico di Padova, Vicolo dell'Osservatorio 5, I--35122, Padova, Italy}
  \footnotetext[3]{MM  acknowledges financial support from the MERAC Foundation, from INAF through PRIN-SKA, from MIUR through Progetto Premiale 'FIGARO' and 'MITiC', and from the Austrian National Science Foundation through FWF stand-alone grant P31154-N27.}
  \arabicfootnotes

%%% Provide here again for all authors of the chapter, the name and affiliation -- only one author at the time!
  \contributor{Michela Mapelli
    \affiliation{Institut f\"ur Astro- und Teilchenphysik, Universit\"at Innsbruck, Technikerstrasse 25/8, A--6020, Innsbruck, Austria}
    \affiliation{INAF -- Osservatorio Astronomico di Padova, Vicolo dell'Osservatorio 5, I--35122, Padova, Italy}}

%%%% PUT HERE YOUR ABSTRACT
 \begin{abstract}
Stellar binaries are the most important energy reservoir of star clusters. Via three-body encounters, binaries can reverse the core collapse and prevent a star cluster from reaching equipartition. Moreover, binaries are essential for the formation of stellar exotica, such as blue straggler stars, intermediate-mass black holes and massive ($\gtrsim{}30$ M$_\odot$) black hole binaries.
 \end{abstract}

\section{Binaries as sources of energy}
\label{sec:reservoir}
For a long time we have known that binaries are an energy reservoir for star clusters. The internal energy of a binary can be expressed as
\begin{equation}
E_{\rm int}=\frac{1}{2}\,{}\mu{}\,{}v^2-\frac{G\,{}m_1\,{}m_2}{r},
\end{equation} 
where $\mu{}=m_1\,{}m_2/(m_1+m_2)$ is the reduced mass, $m_1$ and $m_2$ are the masses of the primary and secondary component of the binary, $G$ is the gravitational constant, $v$ and $r$ are the relative velocity and the distance between the two components of the binary. 

Part of this energy can be exchanged with single stars if they approach the binary in a close encounter (with minimum distance of the order of few times the orbital separation of the binary). In particular, if the binary transfers a fraction of its internal energy to the single star, then the single star's kinetic energy increases, while the binary shrinks (i.e. its binding energy increases). In this case, we say that the binary {\it hardens}. Vice versa, if the binary acquires kinetic energy from the single star, then its semi-major axis increases and we say that the binary {\it softens}. The binary might even be {\it ionized}, i.e. its binding energy may become zero. 

In a star cluster, we usually define as {\it hard} ({\it soft}) binaries those binaries with binding energy $E_{\rm b}>\frac{1}{2}\langle{}m\rangle{}\sigma{}^2$ ($E_{\rm b}<\frac{1}{2}\langle{}m\rangle{}\sigma{}^2$), where $\langle{}m\rangle{}$ is the average mass of a star in the star cluster and $\sigma{}$ is the velocity dispersion. According to \cite{heggie1975}, which is nothing less than the Bible of three-body encounters, hard binaries tend to harden as a consequence of three-body encounters, while soft binaries tend to soften. This statement has been commonly referred to as the Heggie's law. Under the impulse approximation, the hardening rate of a binary does not depend on the mass of the binary and can be expressed as
\begin{equation}\label{eq:hardening}
\frac{d}{dt}\left(\frac{1}{a}\right)=2\,{}\pi{}\,{}\xi{}\,{}\frac{G\,{}\rho{}}{\sigma{}},
\end{equation}
where $a$ is the semi-major axis of the binary, $\xi{}\sim{}0.2-3$ is a dimensionless efficiency parameter \citep{hills1983,quinlan1996,miller2002}, $\rho{}$ is the local mass density of stars, and $\sigma{}$ is the local velocity dispersion. Based on equation~\ref{eq:hardening}, we can say that a hard binary hardens at a constant rate, depending only on the environment (i.e. the local stellar density and velocity dispersion). However, a slight dependence of the hardening rate on the properties of the binary is hidden in the $\xi{}$ parameter, which can be calibrated only from simulations.

{\it Exchanges} are another possible outcome of three-body encounters. During exchanges, one of the members of the binary is kicked off the binary and replaced by the intruder. Exchanges are most likely to happen when the intruder mass is larger than the mass of one of the members of the binary, leading to the formation of more and more massive binaries \citep{hills1980}.

%%%%%%%%%%%%%%%%%%%%%%%%%%%%%%%%%%FIGURE%%%%%%%%%%%%%%%%%%%%%%%%%%%%%%%%%%%%%%
  \begin{figure}
    \includegraphics[width=6cm]{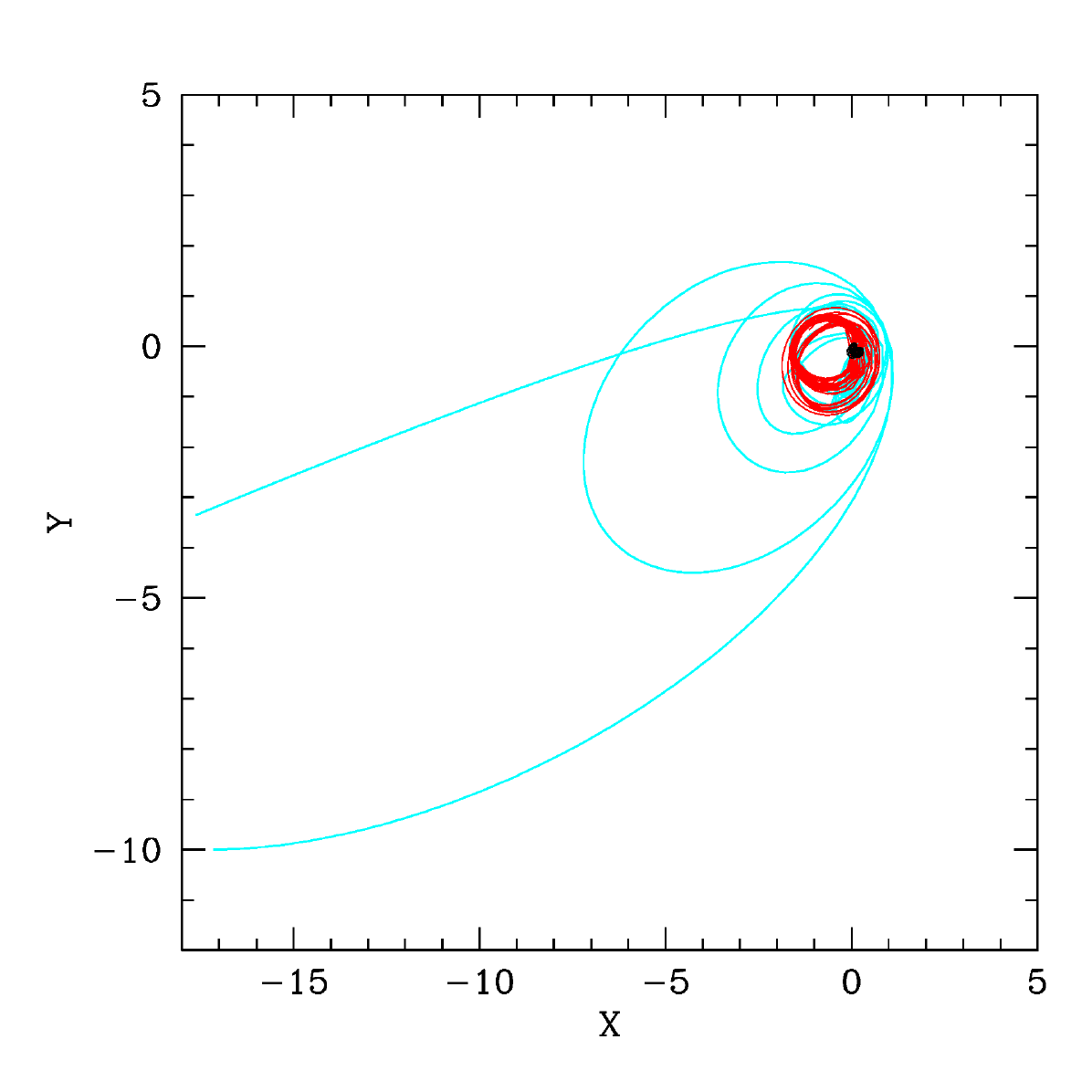}
    \includegraphics[width=6cm]{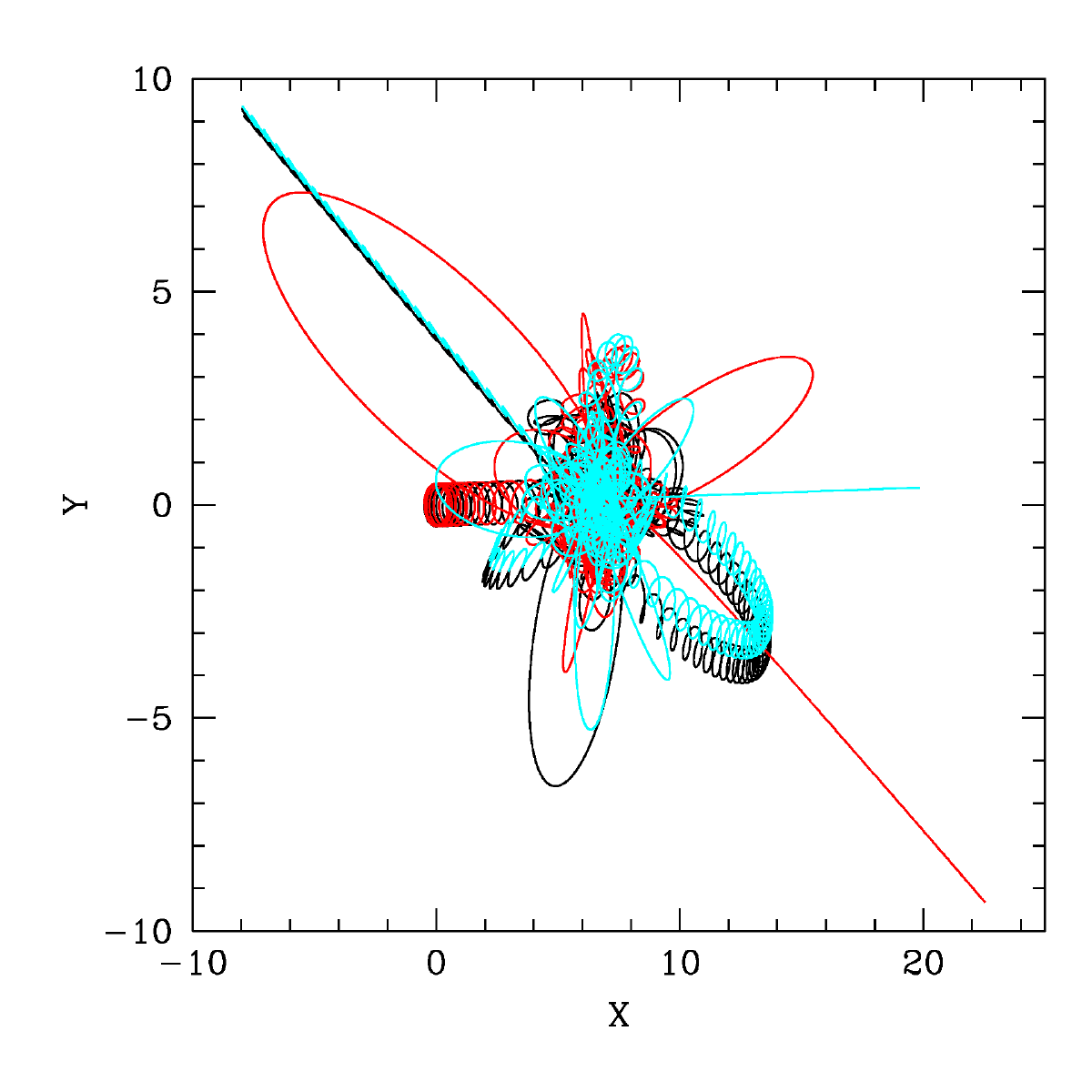}
    \includegraphics[width=6cm]{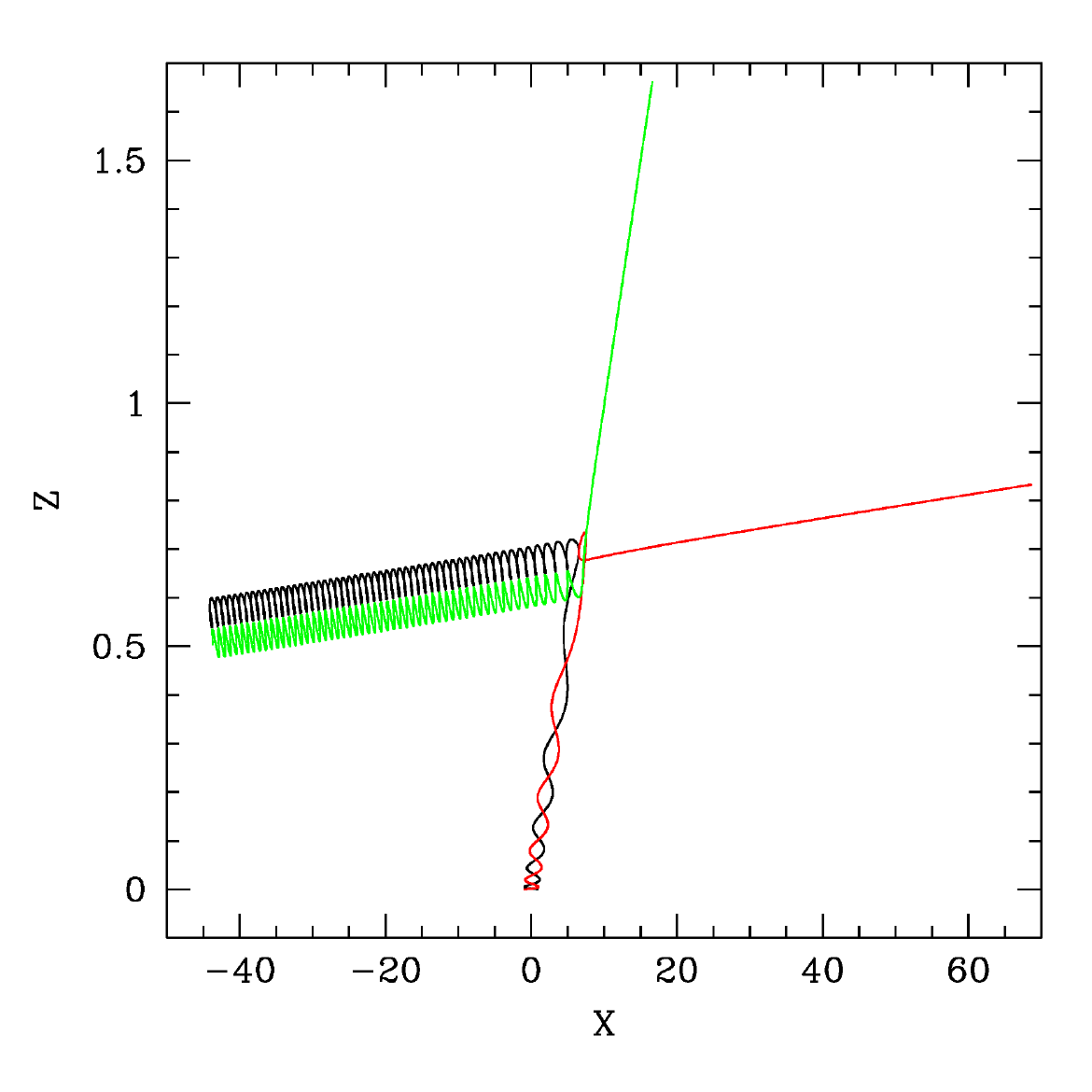}
    \includegraphics[width=6cm]{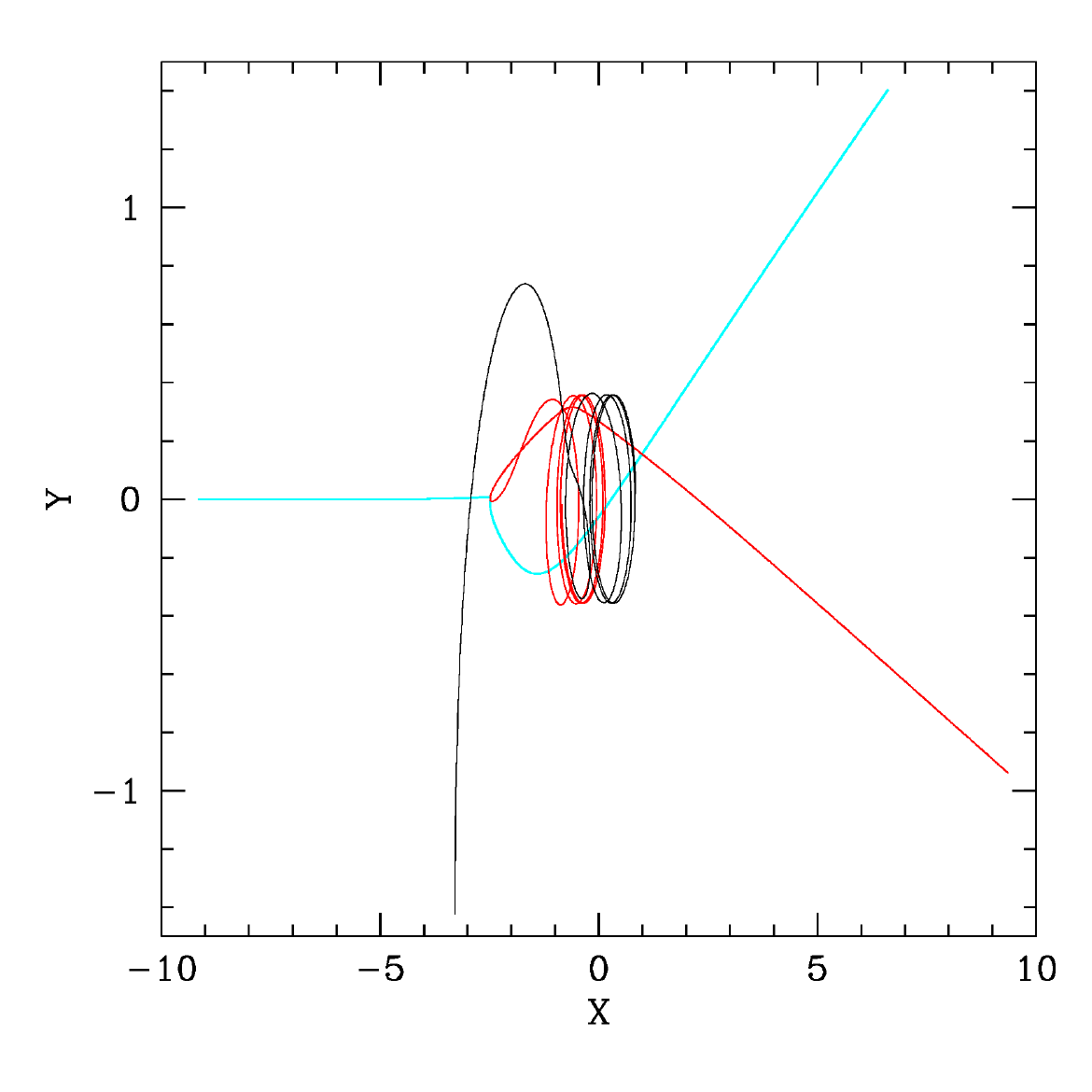}
    \caption[]
      {Simulations of three-body encounters. Top-left: dynamical flyby; top-right: resonant exchange; bottom-left: prompt exchange; bottom-right: ionization. Axes are in $N-$body units.}
    \label{fig:fig0}
     \end{figure}
%%%%%%%%%%%%%%%%%%%%%%%%%%%%%%%%%%FIGURE%%%%%%%%%%%%%%%%%%%%%%%%%%%%%%%%%%%%%%

Figure~\ref{fig:fig0} shows some examples of simulated three-body encounters: a {\it fly-by} (during which energy is exchanged between the binary and the single stars but the binary does not change its components), two exchanges (during which the binary exchanges one of its members with the intruder), and a ionization (during which the binary breaks).
%In this sense, binaries can be considered as a ``Maxwell's demon'' of star clusters, because they can pump kinetic energy  in the system via three-body encounters, changing completely the dynamical state of the system, like Maxwell's demon changes the entropy state of a system in statistical mechanics.

Via three-body encounters, binaries can pump kinetic energy (from their internal energy) in the star cluster, changing its dynamical state completely. In this sense, binaries act like the Maxwell's demon \citep{maxwell1872}, which can change the entropy state of a system in statistical mechanics without any external energy input. The main difference between Maxwell's demons and binaries is that binary systems are definitely real and even rather common astrophysical systems, while Maxwell's demons are just a Gedankenexperiment.

\section{Impact of binaries on core collapse}
\label{sec:corecollapse}
One of the most important and well studied effects of three-body encounters in star clusters is the reverse of core collapse. 

Isolated star clusters composed of a dense core surrounded by a low-density halo are expected to undergo core collapse, because of their negative heat capacity \citep{spitzer1987}. In other words, two-body relaxation is particularly efficient in the core of star clusters, leading to the evaporation of the fastest stars from the core. This leads to a decrease of both the kinetic energy ($K$) and the modulus of the potential energy ($|W|$) of the cluster core, but the decrease of $K$ is more significant than the decrease of $|W|$ because only the fastest stars leave the core. Thus, the core contracts (because potential energy is no longer balanced by kinetic energy), decreasing its radius and increasing its density. If the density increases, two-body encounters are even more efficient (because the two-body relaxation timescale scales as $t_{\rm rlx}\sim{}\rho{}^{-1}\,{}\sigma{}^{3}$, where $\rho$ is the density and $\sigma{}$ the velocity dispersion, \citealt{spitzer1971}), leading to an even faster rate of evaporations. This induces a further contraction, the core radius decreases and the density increases even more, enhancing evaporation, and so on. 

This runaway mechanism, called core collapse, would lead to an infinite core density and a null core radius. Something is needed to break this loop and reverse core collapse, since we do not observe star clusters with pathologically high central densities. A Maxwell's demon is needed to break the loop: something which changes the energetic state of the star cluster.

\begin{figure}
    \includegraphics[width=12cm]{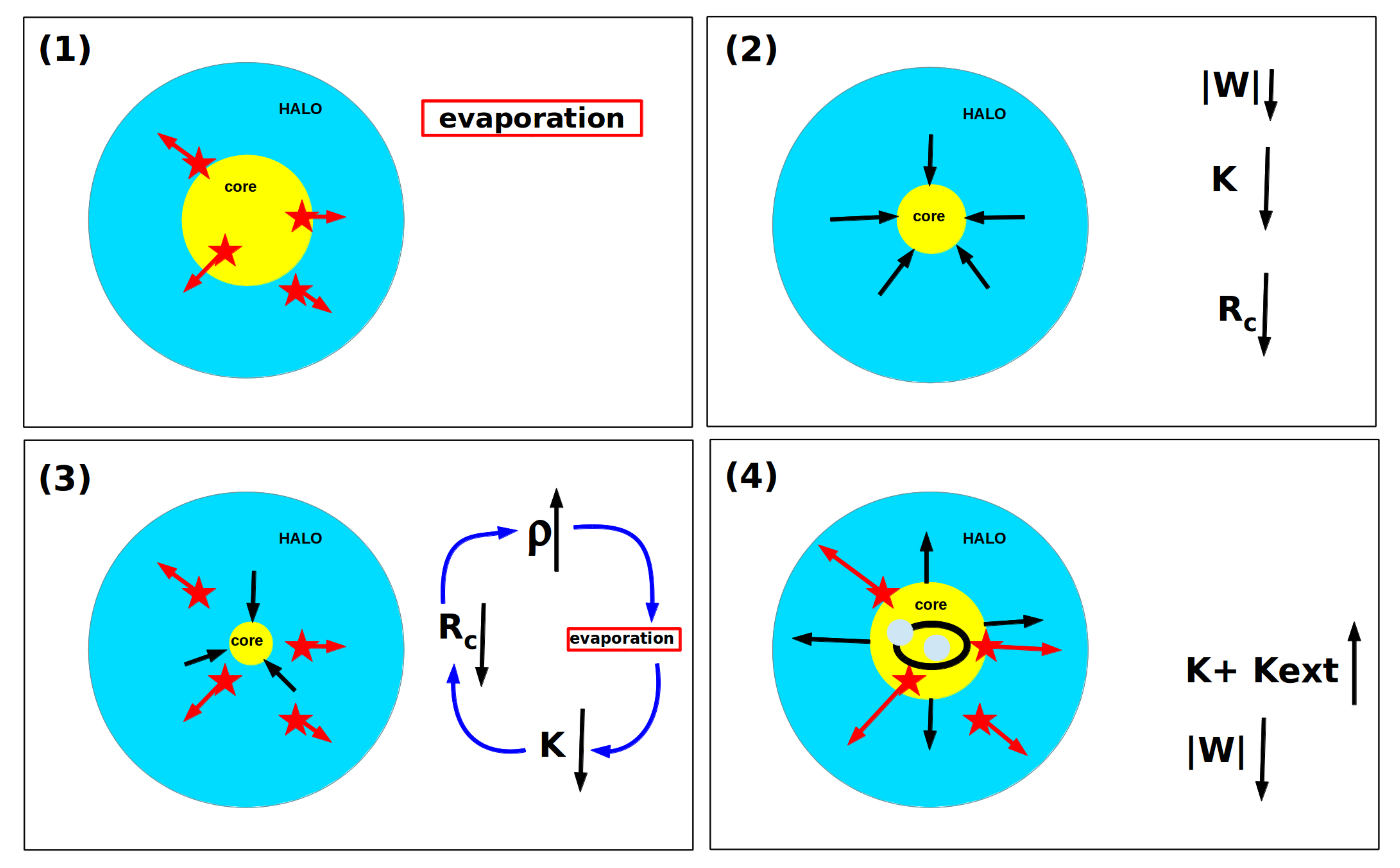}
    \caption[]
      {Cartoon of a core collapse and its reverse. The star cluster is represented as a dense core (yellow) embedded into a low-density halo (blue). In panel (1), stars evaporate from the core because of two-body encounters. The evaporation removes both kinetic energy ($K$) and potential energy ($W$) from the core. Since the fastest stars are more likely to evaporate, $K$ decreases faster than $|W|$: the core contracts (2), because potential energy is no longer balanced by kinetic energy, and the radius of the core $R_{\rm c}$ decreases. The decrease of $R_{\rm c}$ increases the density $\rho{}$ of the core. Thus, two-body encounters become more likely and lead to more efficient evaporation. This further reduces $K$ and the core contracts again, increasing its density (3). The result is a loop (or runaway process): the core keeps contracting forever unless a new source of kinetic energy is switched on in the star cluster. The source of kinetic energy is represented by hard binaries (4), which release kinetic energy ($K_{\rm ext}$) into the core by means of three-body encounters. As an effect of kinetic energy injection, the core expands and the core collapse is reversed.}
    \label{fig:fig1}
     \end{figure}

There are essentially two ways of reversing core collapse: by removing potential energy $|W|$ from the core without removing much kinetic energy and by injecting fresh kinetic energy into the core. \cite{henon1961} suggested for the first time that binaries might be the Maxwell's demon of the situation, able to change the dynamical state of a cluster (see also \cite{elson1987} for a review). When the core density is low, binaries do not exchange significant energy with the other stars, because three-body encounters are rare. During core collapse, when the core density rises by orders of magnitude, three-body encounters become more and more frequent. Moreover, additional binaries can form dynamically during core collapse, because of the higher stellar density. Hard binaries harden, transferring kinetic energy to the other stars. If the intruder stars are not ejected, they increase the kinetic energy of the core. If they are ejected, their kinetic energy is lost but does not trigger a further collapse because these stars were not necessarily the fastest ones before interacting with the binary. The net result is that the kinetic energy of the core increases and the core expands. Figure~\ref{fig:fig1} is a schematic representation of core collapse reversal by binaries.

%%%%%%%%%%%%%%%%%%%%%%%%%%%%%%%%%%FIGURE%%%%%%%%%%%%%%%%%%%%%%%%%%%%%%%%%%%%%%
%%%%%%%%%%%%%%%%%%%%%%%%%%%%%%%%%%FIGURE%%%%%%%%%%%%%%%%%%%%%%%%%%%%%%%%%%%%%%

Several studies \citep{angeletti1977,angeletti1980,chernoff1990,schulman2012,downing2012} suggest that mass loss by massive stars could be another mechanism able to reverse the core collapse, by removing a large fraction of core's potential energy. However, \cite{mapellibressan2013} and \cite{trani2014} highlight that mass loss of massive stars is a viable mechanism to reverse core collapse only if the collapse happens within the lifespan of massive stars, which might be the case only for dense young massive star clusters. In fact, the lifetime of a $\sim{}30$ M$_\odot$ star is $\sim{}6$ Myr, and core collapse happens on a shorter timescale only in star clusters with two-body relaxation timescale $t_{\rm rlx}\lesssim{}50\,{}{\rm Myr}\,{}(M/10^5 M_\odot)^{1/2}\,{}(R/1\,{}{\rm pc})^{3/2}$, assuming that core collapse occurs at $t_{\rm rlx}\sim{}0.1-0.2\,{}t_{\rm rlx}$ (see \citealt{fujii2014}).

%%%%%%%%%%%%%%%%%%%%%%%%%%%%%%%%%%FIGURE%%%%%%%%%%%%%%%%%%%%%%%%%%%%%%%%%%%%%%
  \begin{figure}
    \includegraphics[width=6cm]{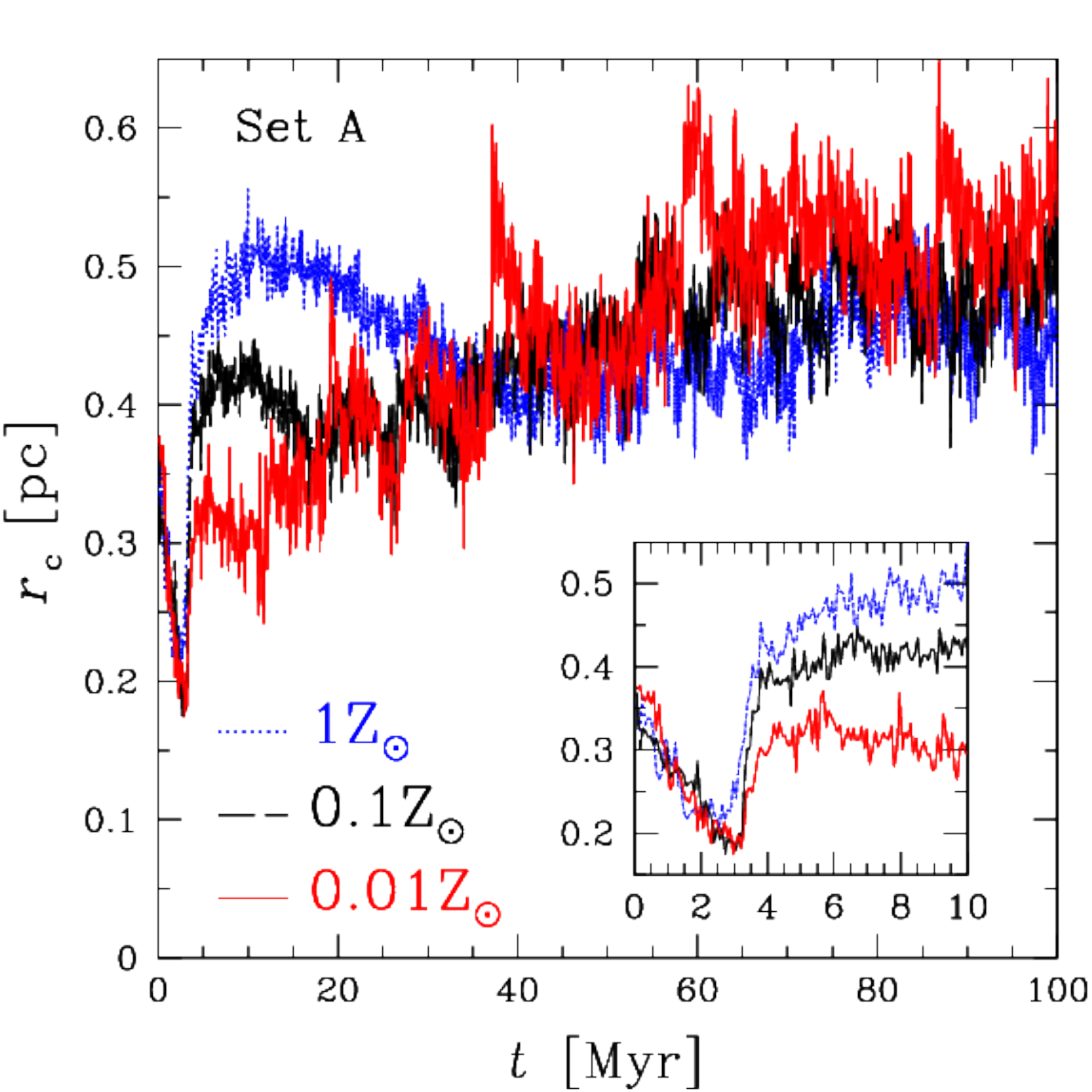}
    \includegraphics[width=6cm]{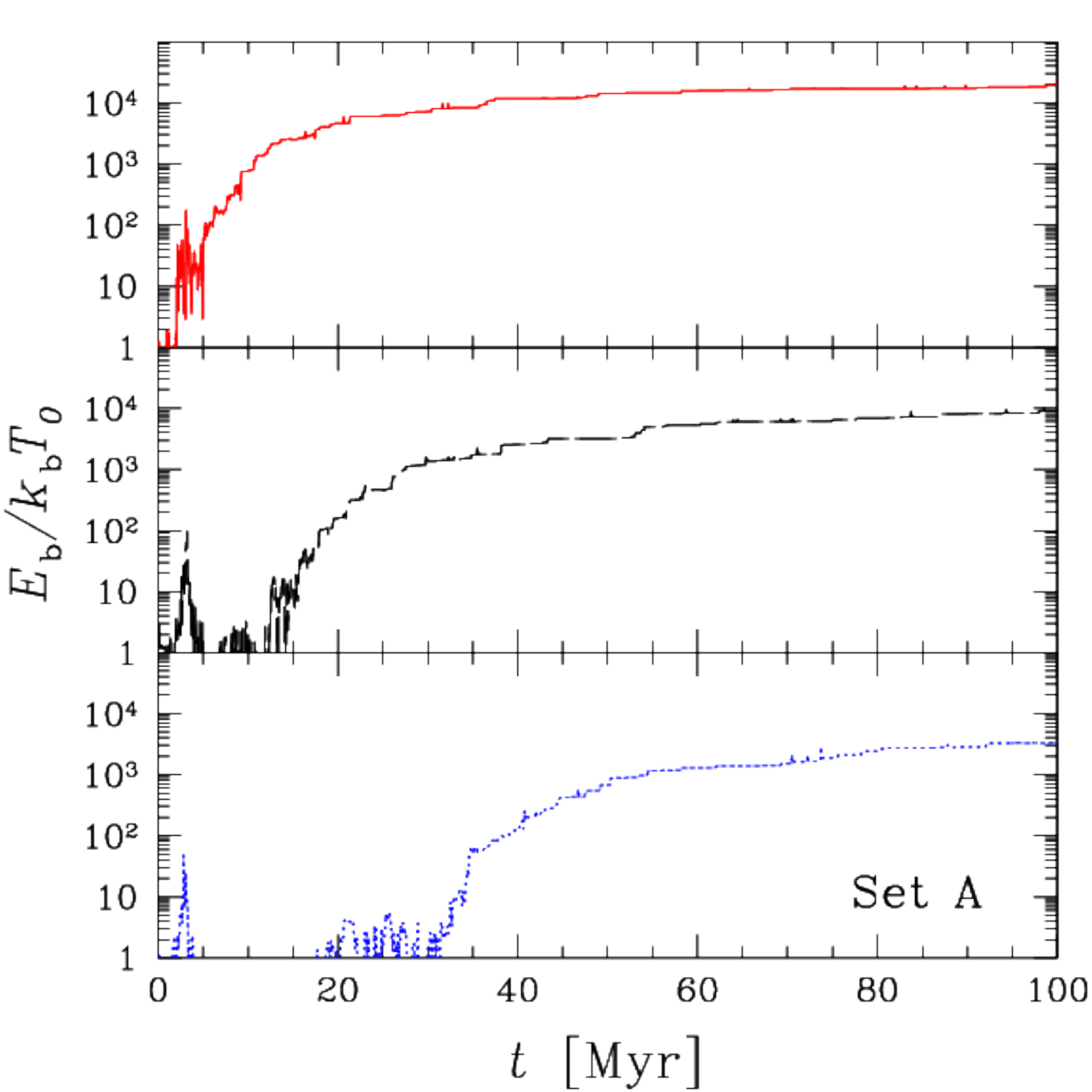}
    \caption[]
            {Core radius (left-hand panel) as a function of time for simulated star clusters with three different metallicities $Z=1, 0.1, 0.01\,{}{\rm Z}_\odot$. In the inset: zoom of the first 10 Myr. Right-hand panel: total internal energy of the binary content of the simulated star clusters as a function of time, normalised to the initial $k_{\rm b}\,{} T_0 = 1/3\,{}\langle{}K\rangle{}|_{t=0}$ , where $\langle{}K\rangle{}$ is the average kinetic energy of a star, $k_{\rm b}$ is the Boltzmann constant, and $T_0$ the initial average kinetic temperature of the star cluster. Solid red line: $Z = 0.01$ Z$_\odot$ ; dashed black
line: $Z = 0.1$ Z$_\odot$ ; dotted blue line: $Z = 1$ Z$_\odot$. Each line is the median value of 10 $N-$body realizations of star clusters, modelled as King models \citep{king1966} with initial virial radius $r_{\rm vir} = 1$ pc and dimensionless central potential  $W_0 = 5$. These simulations start without primordial binaries: all binaries form from dynamical interactions. Adapted from Figure~3 of \cite{trani2014}.
}
    \label{fig:fig2}
     \end{figure}
%%%%%%%%%%%%%%%%%%%%%%%%%%%%%%%%%%FIGURE%%%%%%%%%%%%%%%%%%%%%%%%%%%%%%%%%%%%%%

The left-hand panel of Figure~\ref{fig:fig2} shows the evolution of the core radius in simulations of young dense star clusters with different metallicity. The collapse is apparent at $t\sim{}3$ Myr, followed by a fast reverse. In metal-rich clusters, mass loss by stellar winds is particularly efficient in reversing core collapse, as it is apparent from the right-hand panel of Figure~\ref{fig:fig2}. This figure shows the total binding energy stored in binaries as a function of time. While at low metallicity the binding energy of binaries grows rapidly during and after core collapse, at solar metallicity we must wait for $t\sim{}30$ Myr before the first hard binaries appear. This difference can be easily explained because mass loss by stellar winds depends on metallicity as $\dot{M}\propto{}Z^\alpha{}$ with $\alpha{}\sim{}0.85$ \citep{vink2001}.

\section{Impact of binaries on equipartition}
\label{sec:equipartition}
Core collapse reverse is the best known effect of binaries on the global evolution of star clusters, but there are also other crucial, although more subtle effects, which are still matter of investigation. 

For example, hard binaries might play a crucial role in the onset of Spitzer's instability. From the equipartition theorem of statistical mechanics \citep{boltzmann1876}, we know that in gas systems at thermal equilibrium energy is shared equally by all particles. For analogy with gas, we expect that in a two-body relaxed stellar system the kinetic energy of a star $i$ is {\emph locally} the same as that of the star $j$ ($m_i\,{}v_i^2\sim{}m_j\,{}v_j^2$). Thus, if a stellar system is in equipartition, the local velocity of a star\footnote{It is important to stress that the equipartition theorem is a local theorem: it does not apply to the star cluster as a whole.} should depend on its mass as $v(m)\propto{}m^{-0.5}$. If the velocities of all stars are initially drawn from the same distribution, massive stars are then expected to transfer kinetic energy to lighter stars and slow down, till they reach equipartition.

\cite{spitzer1969} already predicted that there are star clusters which cannot reach equipartition. According to Spitzer's analytic model, an idealized star cluster composed of two classes of stars (the stars in the first and the second class having mass $m_1$ and $m_2$, respectively, with $m_1<<m_2$) cannot reach equipartition if $M_2>0.16\,{}M_1\,{}(m_2/m_1)^{3/2}$, where $M_1$ and $M_2$ are the total mass of class 1 and 2, respectively. This has a concrete physical interpretation: if there are too many massive stars in a star cluster, then the light stars are not able to absorb all the excess of kinetic energy that the massive stars need to transfer to them in order to reach equipartition. The light stars are ejected on larger orbits, while the massive stars decouple dynamically from the light stars and tend to interact between each other, forming binary systems and eventually ejecting each other through three-body encounters \citep{bonnelldavies1998,allison2009,portegieszwart2010}. This process takes the name of Spitzer's instability.

While the onset of Spitzer's instability has been demonstrated analytically only for rather idealized systems \citep{spitzer1969,vishniac1978,merritt1981,miocchi2006}, Monte Carlo \citep{bianchini2016} and N-body simulations \citep{trenti2013,parker2016,spera2016} show that Spitzer's instability is rather common, if not universal in star clusters (both open clusters and globular clusters).

%%%%%%%%%%%%%%%%%%%%%%%%%%%%%%%%%%FIGURE%%%%%%%%%%%%%%%%%%%%%%%%%%%%%%%%%%%%%%
  \begin{figure}
    \includegraphics[width=13cm]{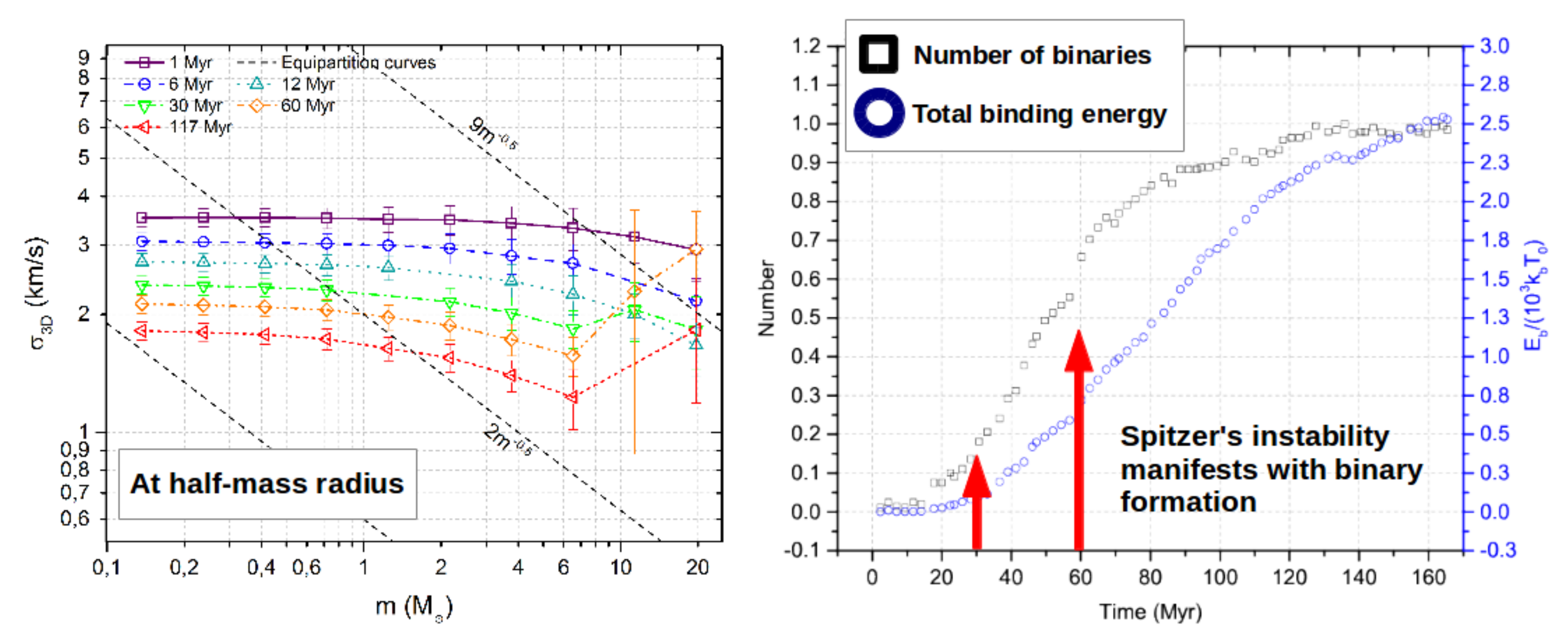}
    \caption[]
      {{\emph Left-hand panel}: three-dimensional velocity dispersion $\sigma{}_{\rm 3D}$ as a function of the stellar mass in 200 realizations of star clusters with $N=10^4$ and two-body relaxation timescale $t\sim{}27$ Myr. The lines show the median values of $\sigma{}_{\rm 3D}$, the error bars represent the standard deviation over 200 different Monte Carlo realizations. Different lines correspond to a different time since the beginning of the simulation. Solid purple line with open squares: $t=1$ Myr; dashed blue lines with open circles: $t=6$ Myr; dotted turquoise line with triangles: $t=12$ Myr; dash-dot green line with downward triangles: $t=30$ Myr; dash-double dot orange line with diamonds: $t=60$ Myr; short dashed red line with leftward open triangles: $t=117$ Myr. Dashed black diagonal lines correspond to the trend expected from equipartition ($\sigma{}_{\rm 3D}\propto{}m^{-0.5}$). Only stars inside half-mass radius are considered. {\emph Right-hand panel}: total number of binaries (open black squares and left $y$ axis) and total binding energy (open blue circles and right $y$ axis) as a function of time in the same simulations as shown in the left-hand panel. Vertical red arrows correspond to $t=30$ and $t=60$ Myr. These simulations start without primordial binaries: all binaries form from dynamical interactions. Adapted from Figures~4 and 11 of \cite{spera2016}.}
    \label{fig:fig3}
     \end{figure}
%%%%%%%%%%%%%%%%%%%%%%%%%%%%%%%%%%FIGURE%%%%%%%%%%%%%%%%%%%%%%%%%%%%%%%%%%%%%%

The left-hand panel of Figure~\ref{fig:fig3} is an example of the onset of Spitzer's instability in star clusters. The massive stars initially transfer kinetic energy to the light stars and their velocity dispersion decreases, approaching the trend expected for equipartition. However, at time $\gtrsim{}30$ Myr massive stars become dynamically hotter than lower mass stars, and the process of reaching equipartition is interrupted: Spitzer's instability develops. The right-hand panel of Figure~\ref{fig:fig3} shows that there is an interesting correlation between the formation of binary systems and the onset of Spitzer's instability. The simulations shown in Fig.~\ref{fig:fig3} start from zero binary systems. Thus, all binaries in the clusters have dynamical origin. It is apparent that the formation of binaries starts exactly at $\sim{}30$ Myr, when Spitzer instability also manifests itself, and then the number of binaries and their total binding energy grow very fast. A possible explanation of this link between binary formation/hardening and the onset of Spitzer's instability is the following. Spitzer's instability implies that massive stars decouple dynamically from light stars; thus massive stars can interact only between each other, forming binaries and undergoing three body encounters which increase their velocity dispersion and make them hotter. %and eject some of them from the core.

\section{Stellar Exotica}
\label{sec:exotica}
The dynamics of binaries in star clusters is essential to generate the so called {\it stellar exotica}, i.e. stellar-like objects that cannot be produced (or can hardly be produced) by stellar evolution alone \citep{davies2002}. Famous examples of stellar exotica are blue straggler stars (BSSs) and intermediate-mass black holes (IMBHs, i.e. black holes with mass $\sim{}10^2-10^5$ M$_\odot$), but even massive stellar black hole (BH) binaries, such as the ones observed by the LIGO-Virgo collaboration \citep{abbott2016a}, might be part of this category.
\subsection{Blue straggler stars}
Lying on the continuation of the main sequence, but above the turn-off in the colour-magnitude diagram of a star cluster, BSSs appear to be core Hydrogen burning stars which are more massive than the turn-off. This has been interpreted as the effect of stellar rejuvenation, which might come from mass transfer in a binary or from the collision (and merger) between two main sequence stars, likely triggered by a three-body encounter. In both cases, a binary is needed to form the BSS. %BSSs require a binary system to form. 

Since their formation involves the dynamics of binaries, BSSs are thought to be a good indicator of the dynamical evolution of a star cluster. In these proceedings, Ferraro et al. (2018, see also \citealt{mapelli2004,mapelli2006,beccari2006}) highlight the fact that the radial distribution of BSSs, normalized to the average light distribution in a star cluster, is suggestive of the dynamical evolution of a star cluster: BSSs are among the most massive objects in a star cluster; thus, they sink to the centre by dynamical friction. The presence of a minimum in the normalized distribution of BSSs indicates the maximum distance from the centre below which dynamical friction was efficient in sweeping BSSs toward the core. 
%BSSs sank to the core by dynamical friction. 

In this scenario, BSSs would be a powerful indicator of the dynamical friction timescale, and thus also of the two-body relaxation timescale of a star cluster\footnote{The two-body relaxation timescale $t_{\rm rlx}$ and the dynamical friction timescale $t_{\rm DF}(M)$ of a star with mass $M$ are linked by the following relation: $t_{\rm DF}(M)=\frac{\langle{}m\rangle{}}{M}\,{}t_{\rm rlx}$, where $\langle{}m\rangle{}$ is the average mass of a star in the star cluster.}. In this sense, BSSs would be a genuine indicator of the dynamical age of a star cluster: a `dynamical clock', as coined by \cite{ferraro2012}.

Have we completely understood BSSs, then? No. Our picture of BSSs is still far from optimal. For example, \cite{hypki2017} simulate the dynamical evolution of BSSs in a globular cluster through the MOCCA Monte Carlo code. They produce the normalized distribution of BSSs from their simulations as a function of time, finding that the position of the minimum and the shape of the curve shift much faster than a two-body relaxation timescale. This suggests that we still do not have a complete understanding of BSSs and we should keep investigating on them, from both the theoretical and the observational point of view (see both Francesco Ferraro's and Robert Mathieu's contribution to these proceedings).

\subsection{Massive black hole binaries}
On September 14 2015, the two LIGO interferometers captured the first gravitational wave signal from a merging BH binary \citep{abbott2016a}. GW150914 has been interpreted as the merger of two BHs with mass $m_{\rm BH}=36.2^{+5.2}_{-3.8},\,{}29.1^{+3.7}_{-4.4}$ M$_\odot$ \citep{abbott2016b}. Since then, the LIGO-Virgo collaboration has detected five BH mergers, three of them containing BHs more massive than $\sim{}30$ M$_\odot$ \citep{abbott2016b,abbott2017a,abbott2017b,abbott2017c}. 

The formation channel of these massive BH binaries is still an open question. Two main channels have been proposed in the last few years: isolated binary evolution and dynamical formation in star clusters.

In the isolated binary evolution scenario, two massive stars form gravitationally bound, from the same cloud, and then they evolve through several processes (e.g. Roche lobe overflow, common envelope, etc) till they become two BHs and eventually merge. The two progenitor stars should be relatively metal poor ($Z\lesssim{}0.3$ Z$_\odot$) in order to end up as two massive BHs \citep{mapelli2009,belczynski2010,spera2015,belczynski2016,mapelli2017,giacobbo2018}.

Alternatively, BH binaries can form from dynamical processes in star clusters. For example, a dynamical exchange can lead to the formation of a double BH binary in the core of a dense star cluster. \cite{ziosi2014} show that more than $\sim{}90$ per cent of the double BH binaries in star clusters form through dynamical exchanges. BHs are particularly efficient in capturing companions through dynamical exchanges, because they are more massive than the vast majority of the other stars in star clusters. As we mentioned in Section~\ref{sec:reservoir} exchanges are more likely if the mass of the intruder is larger than the mass of one of the members of the binary. Thus, exchanges favour the formation of more and more massive BH binaries, similar to the ones observed by LIGO and Virgo. Moreover, BH binaries produced by exchanges have initially large eccentricities. A third crucial feature of BH binaries formed by exchanges is the distribution of BH spins: exchanges tend to randomize spins, and we expect a nearly isotropic distribution of BH spins in dynamically formed binaries.

\begin{figure}
    \includegraphics[width=10cm]{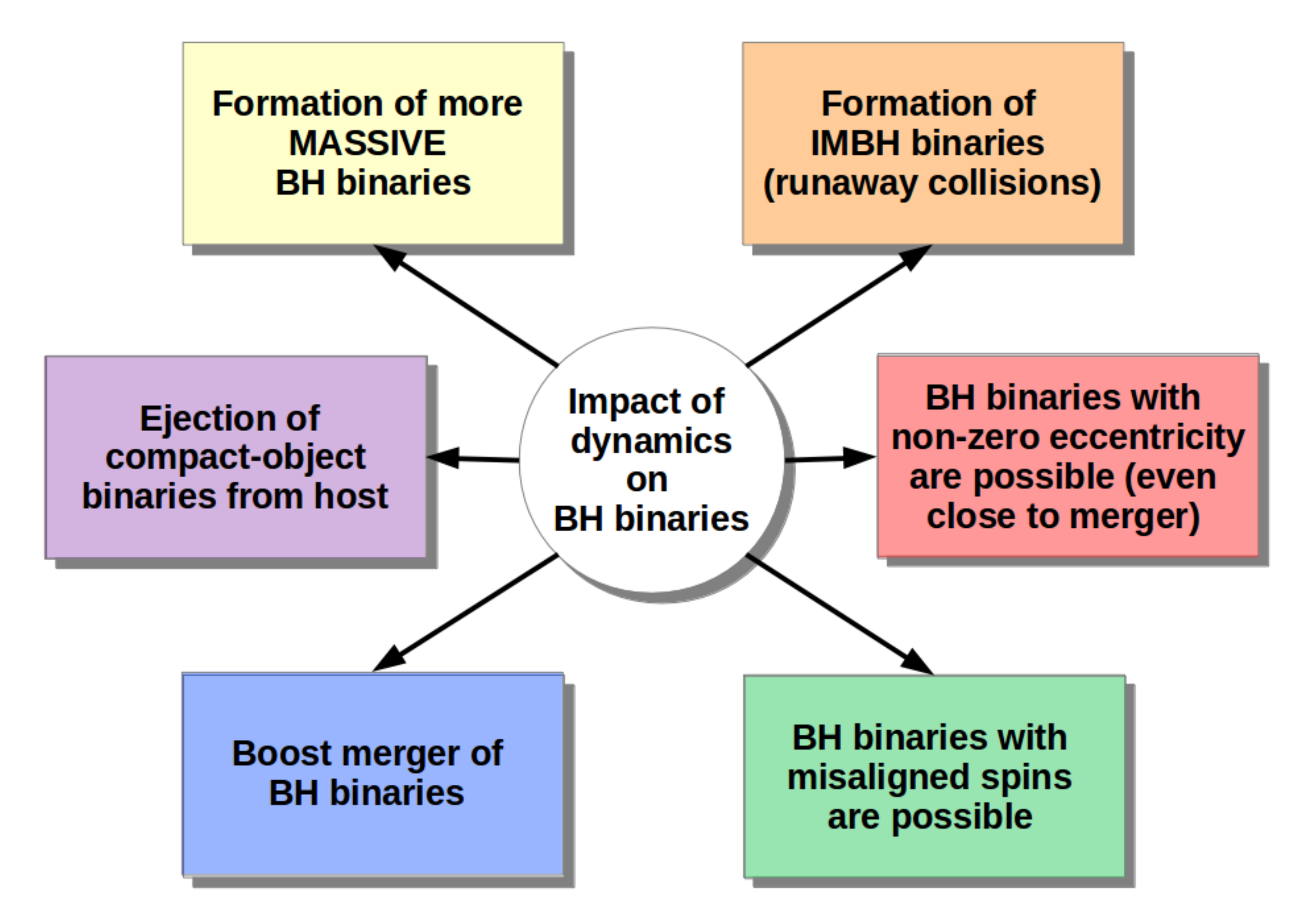}
    \caption[]
      {Summary of the main effects of star cluster dynamics on the formation and evolution of BH binaries.}
    \label{fig:figERC}
     \end{figure}

Moreover, BH binaries are expected to be relatively hard binaries; thus, three-body encounters can harden them even more, possibly enhancing the merger rate of BH binaries. Finally, recoil kicks induced by three-body encounters can also eject BH binaries from their parent star clusters \citep{mapelli2013}. Figure~\ref{fig:figERC} is a summary of the main effects of star cluster dynamics on BH binaries.

%%%%%%%%%%%%%%%%%%%%%%%%%%%%%%%%%%FIGURE%%%%%%%%%%%%%%%%%%%%%%%%%%%%%%%%%%%%%%
%%%%%%%%%%%%%%%%%%%%%%%%%%%%%%%%%%FIGURE%%%%%%%%%%%%%%%%%%%%%%%%%%%%%%%%%%%%%%

Hardening by three-body encounters is thus a crucial mechanism to enhance the merger rate of BH binaries. It is even possible to make a simple analytic estimate of the evolution of the semi-major axis $a$ of a double BH binary which is affected by three-body encounters and by gravitational wave emission (see equation~9 of \citealt{colpi2003}):
\begin{equation}\label{eq:miatesi}
\frac{da}{dt}=-2\,{}\pi{}\,{}\xi{}\,{}\frac{G\,{}\rho{}}{\sigma{}}\,{}a^2-\frac{64}{5}\frac{G^3\,{}m_1\,{}m_2\,{}(m_1+m_2)}{c^5\,{}(1-e^2)^{7/2}}\,{}a^{-3},
\end{equation}
where $\xi{}\sim{}0.2-3$ is the same dimensionless parameter as defined in equation~\ref{eq:hardening}, $\rho{}$ is the local mass density of stars, $\sigma{}$ is the local velocity dispersion, $c$ is the light speed, and $e$ is the eccentricity of the binary. 
The first part of the right-hand term of equation~\ref{eq:miatesi} accounts for the effect of three-body hardening on the semi-major axis (from equation~\ref{eq:hardening}). It scales as $da/dt\propto{}-a^2$, indicating that the larger the binary is, the more effective the hardening. This can be easily understood considering that the geometric cross section for three body interactions with a binary scales as $a^2$.

The second part of the right-hand term of equation~\ref{eq:miatesi} accounts for energy loss by gravitational-wave emission. It is the first-order approximation of the calculation by \cite{peters1964}. It scales as $da/dt\propto{}-a^{-3}$ indicating that gravitational-wave emission becomes efficient only when the two BHs are very close to each other.

%%%%%%%%%%%%%%%%%%%%%%%%%%%%%FIGURE %%%%%%%%%%%%%%%%%%%%%%%%%%%%%%%%%%%%%%%%
\begin{figure}
\center{
\includegraphics[width=9cm]{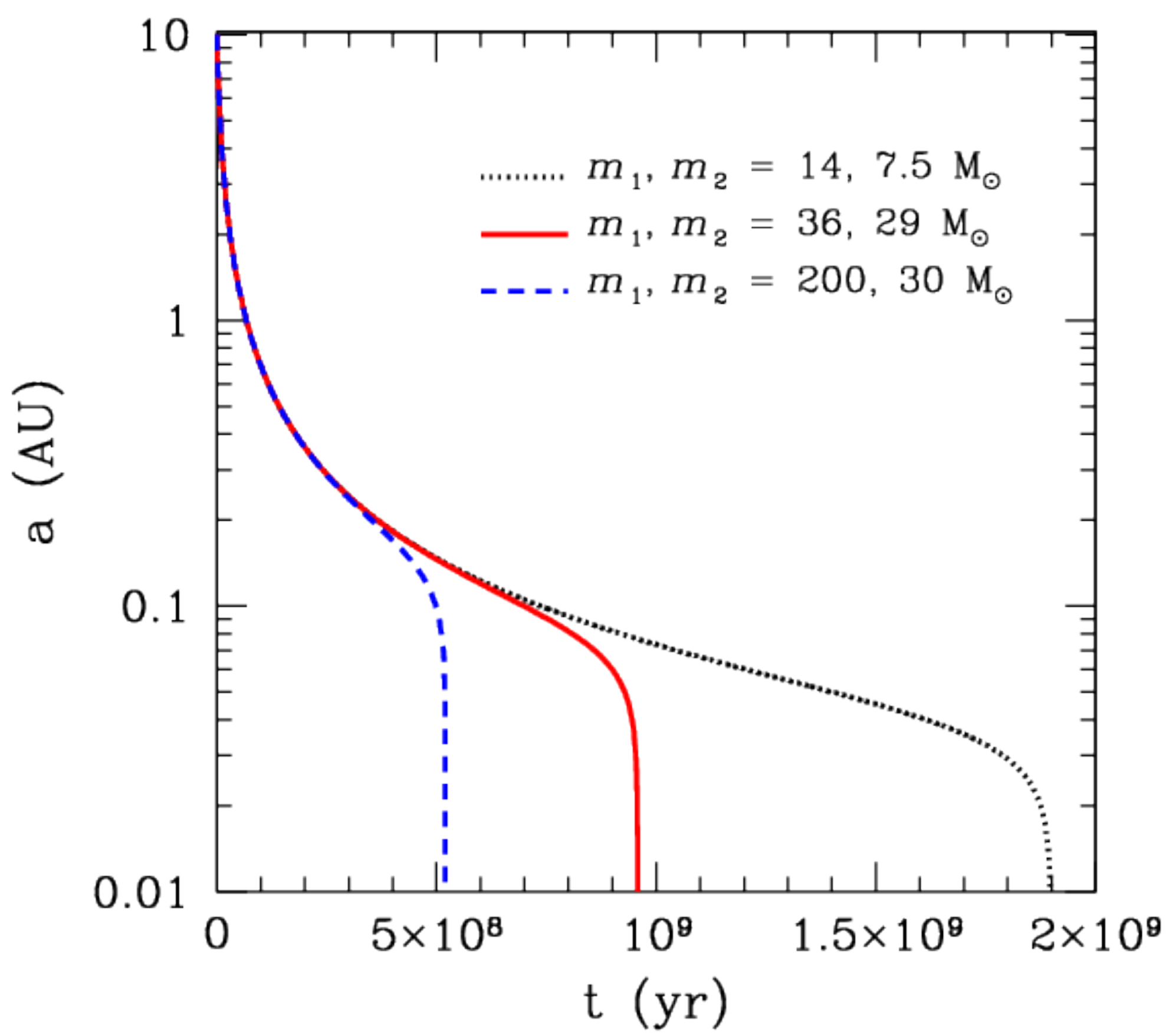}     % includes figure foo.eps
}
\caption{\label{fig:miatesi}Time evolution of the semi-major axis of three BH binaries estimated from equation~\ref{eq:miatesi}. Blue dashed line: BH binary with masses $m_1=200$ M$_\odot$, $m_2=30$ M$_\odot$; red solid line: $m_1=36$ M$_\odot$, $m_2=29$ M$_\odot$; black dotted line: $m_1=14$ M$_\odot$, $m_2=7.5$ M$_\odot$. For all BH binaries: $\xi{}=1$, $\rho{}=10^5$ M$_\odot$ pc$^{-3}$, $\sigma{}=10$ km s$^{-1}$, initial semi-major axis of the BH binary $a_{\rm i}=10$ AU, eccentricity $e=0$.}
\end{figure}
%%%%%%%%%%%%%%%%%%%%%%%%%%%%%%%%%%%%%%%%%%%%%%%%%%%%%%%%%%%%%%%%%%%%%%%%%%%%

In Figure~\ref{fig:miatesi} we solve equation~\ref{eq:miatesi} numerically for three double BH binaries with different mass. All binaries evolve through (i) a first phase in which hardening by three body encounters dominates the evolution of the binary, (ii) a second phase in which the semi-major axis stalls because three-body encounters become less efficient as the semi-major axis shrink, but the binary is still too large for gravitational-wave emission to become efficient, and (iii) a third phase in which the semi-major axis drops because the binary enters the regime where gravitational-wave emission is efficient.

%While the evolution of isolated binaries at low metallicity can explain such massive  
\subsection{Intermediate-mass black holes}
Our understanding of intermediate-mass black holes (IMBHs) is remarkably poor. We are not even sure they exist, because the observational constraints are rather scanty:
\begin{itemize}
\item{}ESO 243-49 HLX-1 is an IMBH candidate \citep{farrell2009,servillat2011} with mass $\sim{}10^4$ M$_\odot$. It is associated with a hyper-luminous X-ray source (peak luminosity $\sim{}10^{42}$ erg s$^{-1}$) whose light curve shows some periodicity and a fast-rise exponential decay behaviour;
\item{}\cite{kiziltan2017} claim that the motion of pulsars is suggestive of a $2300_{-850}^{+1500}$ M$_\odot$ IMBH in the centre of 47~Tucanae;  
\item{}other IMBH candidates are supported by the combination of spectra and photometry in globular clusters (e.g. G1, \citealt{gebhardt2005});
\item{}LIGO-Virgo provide an upper limit to the IMBH merger rate of $R_{\rm IMBH}<0.93$ Gpc$^{-3}$ yr$^{-1}$ in the range $100-500$ M$_\odot$ \citep{abbott2017IMBH}.
\end{itemize}

From the theoretical point of view, the formation channels of IMBHs are deeply connected with the dynamics of binaries. Here below, we will describe the repeated merger and the runaway collision scenario, which are two of the most popular theoretical models for the formation of IMBHs.

\begin{figure}
    \includegraphics[width=10cm]{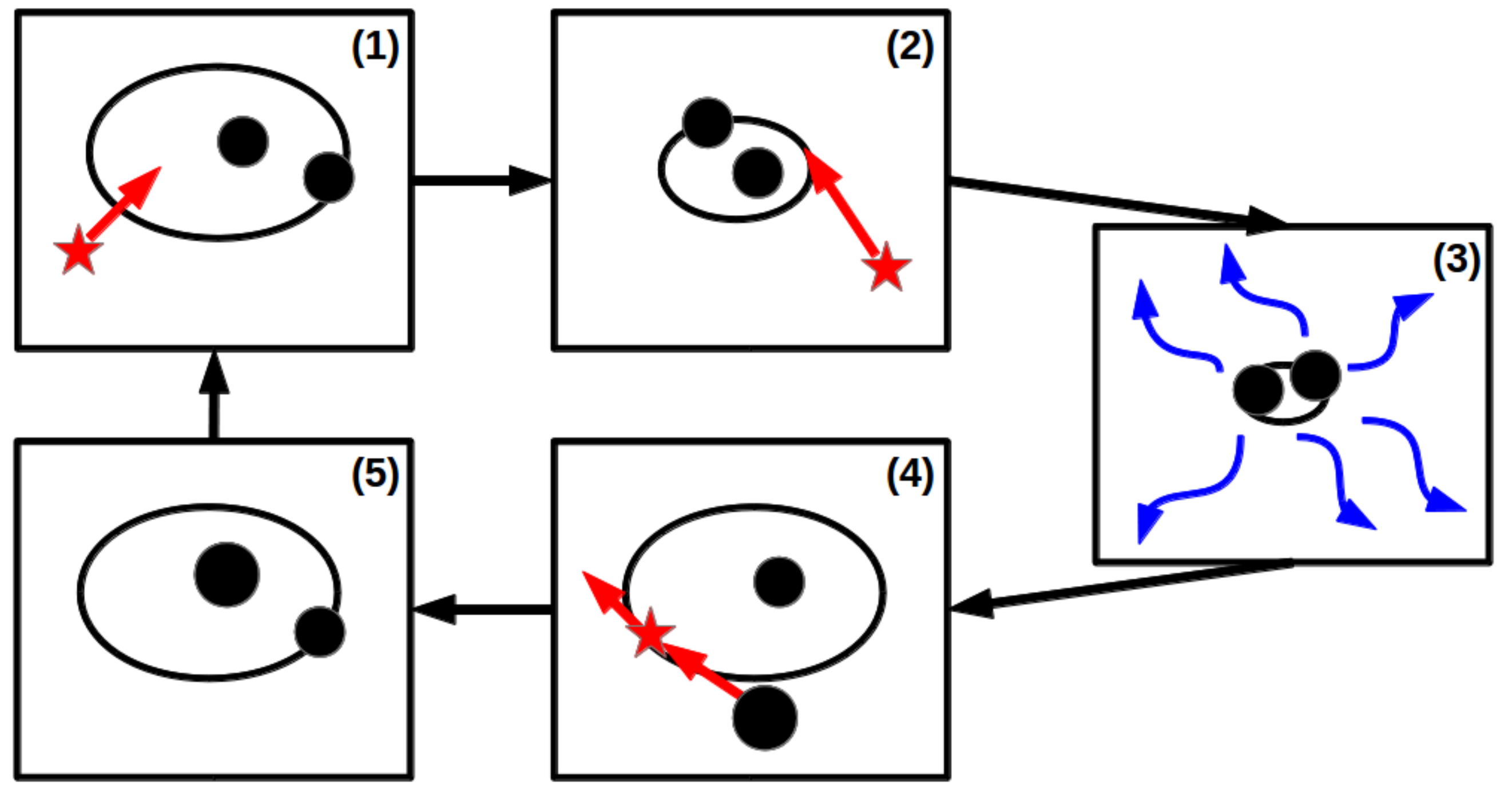}
    \caption[]
{Cartoon of the repeated merger scenario in old star clusters (see e.g. \citealt{miller2002, giersz2015}). From top to bottom and from left to right: (1) a BH binary undergoes three-body encounters in a star cluster; (2) three-body encounters
harden the BH binary, shrinking its semi-major axis; (3) the BH binary hardens by three-body encounters till it enters the regime where gravitational wave emission is efficient: the binary semi-major axis decays by gravitational wave emission and the binary merges; (4) a single bigger BH forms as result of the merger, which may acquire a new companion by dynamical exchange (if it is not ejected by gravitational wave recoil); (5) the new binary containing the bigger BH starts shrinking again by three-body encounters (1). This loop may be repeated several times till the main BH becomes an IMBH.
}
    \label{fig:repeated}
     \end{figure}

In the {\it repeated merger} scenario \citep{miller2002,giersz2015}, a stellar-mass BH binary in the core of a globular cluster undergoes repeated three-body encounters, hardening till  it can merge by gravitational wave emission (if it is not ejected by a dynamical kick). This leads to the formation of a more massive BH, which, if it is not ejected by gravitational-wave recoil, can enter a new binary by exchange. The new binary hardens by three-body encounters till it can merge by gravitational-wave emission, and so on (see Fig.~\ref{fig:repeated}).

%%%%%%%%%%%%%%%%%%%%%%%%%%%%%%%%%%FIGURE%%%%%%%%%%%%%%%%%%%%%%%%%%%%%%%%%%%%%%
%%%%%%%%%%%%%%%%%%%%%%%%%%%%%%%%%%FIGURE%%%%%%%%%%%%%%%%%%%%%%%%%%%%%%%%%%%%%%

The main issue of the repeated merger scenario is that BHs might be ejected by three-body kicks or by gravitational-wave recoils before accreting enough mass to become IMBHs. Moreover, the assembly by repeated mergers is rather inefficient.

\begin{figure}
    \includegraphics[width=12cm]{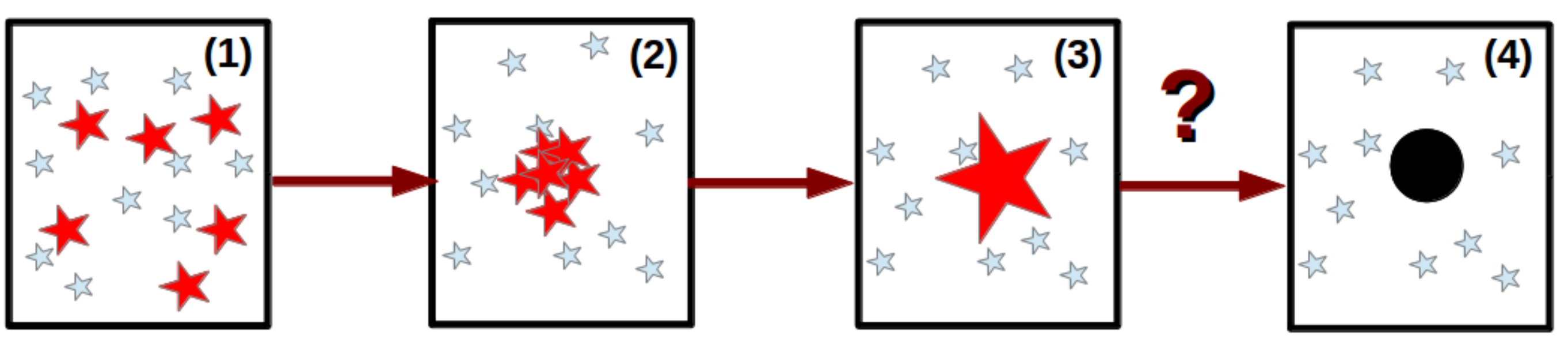}
    \caption[]
{Cartoon of the runaway collision scenario in dense young star clusters (see e.g. \citealt{portegieszwart2002}). From left to right: (1) the massive stars (red big stars) and the low-mass stars (blue small stars) follow the same initial spatial distribution; (2) dynamical friction leads the massive stars to sink to the core of the cluster, where they start colliding between each other; (3) a very massive star ($\gg{}100$ M$_\odot$) forms as a consequence of the runaway collisions; (4) this massive star might be able to directly collapse into a BH.
}
    \label{fig:runaway}
     \end{figure}

In contrast, the {\it runaway collision} scenario is expected to happen only in very young massive dense star clusters (such as R136 in the Large Magellanic Cloud), where the dynamical friction timescale for very massive stars is shorter than the time for the first supernovae to explode \citep{portegieszwart2002,portegieszwart2004}. If this condition is satisfied, the most massive stars sink to the centre of the star cluster before they explode as supernovae, enhancing the central density of massive objects. Once in the core, the massive stars are likely to form binaries and to collide with each other by three-body encounters. This leads to the formation of a very massive star, as shown in Fig.~\ref{fig:runaway}. The fate of the massive star represents the main uncertainty of the runaway collision scenario. In fact, the star might either collapse to an IMBH directly or produce a smaller remnant via a supernova explosion.

Recent studies \citep{mapelli2016,spera2017} suggest that the formation of an IMBH is inhibited at solar metallicity, because of the strong stellar winds. At low metallicity ($\lesssim{}0.1$ Z$_\odot$), about $10-20$ per cent of the runaway collision products might produce IMBHs with mass $\sim{}100-500$ M$_\odot$. Most runaway collision products are expected to form double BH binaries by exchanges, with important implications for gravitational wave detections \citep{mapelli2016}.

%%%%%%%%%%%%%%%%%%%%%%%%%%%%%%%%%%FIGURE%%%%%%%%%%%%%%%%%%%%%%%%%%%%%%%%%%%%%%
%%%%%%%%%%%%%%%%%%%%%%%%%%%%%%%%%%FIGURE%%%%%%%%%%%%%%%%%%%%%%%%%%%%%%%%%%%%%%

\section{Summary}
\label{sec:summary}

We reviewed the importance of binaries for the dynamical evolution of a star cluster. The internal energy of binaries represents the main energy reservoir of a star cluster. When they interact via three-body and multiple-body encounters, binaries can exchange a fraction of their internal energy with the other stars in the cluster. In particular, hard binaries (i.e. binaries whose binding energy is larger than the average kinetic energy of a star in the star cluster) tend to transfer a part of their kinetic energy to the other stars  \citep{heggie1975}.

From this perspective, binaries act as ``Maxwell's demons'' of star clusters, because they change the dynamical state of a star cluster without any external energy input, just by exchanging their internal energy with stars. Three-body encounters with hard binaries are also called ``super-elastic'' encounters, because they increase the translational kinetic energy of the involved bodies.

The importance of binaries as energy reservoirs is apparent from the study of core collapse: star clusters are subject to gravothermal catastrophe because of their negative heat capacity. This leads their core to contract in a runaway sense. Three-body encounters are the most effective process in reversing core collapse, because they pump fresh kinetic energy into the core (extracted from the internal energy of hard binaries) and stop the collapse (see \citealt{spitzer1987} and references therein).

While the reverse of core collapse is the best known example of the importance of binaries for star clusters, binaries play a crucial role in many other dynamical processes. For example, N-body simulations \citep{spera2016} show that binary hardening is a clear indicator of the onset of Spitzer's instability \citep{spitzer1969}.

Finally, binaries are necessary to explain the formation of stellar exotica,  i.e. stellar-like objects that cannot be produced (or can hardly be produced) by `simple' stellar evolution. For example, blue straggler stars are interpreted as the product of  mass transfer in binaries or of stellar collisions, triggered by three-body encounters. 

Even BH binaries are efficiently produced by three-body encounters, especially by dynamical exchanges \citep{ziosi2014}. Exchanges tend to produce more massive BH binaries than isolated binary evolution, with higher eccentricity and with a nearly isotropic spin distribution. Moreover, dynamical hardening is expected to enhance the merger rate of BH binaries \citep{rodriguez2016,askar2017,banerjee2017}. Finally, dynamical processes such as the runaway collision \citep{portegieszwart2004,mapelli2016} and the repeated merger scenario \citep{miller2002,giersz2015} are among the most popular models to explain the formation of IMBHs. Thus, the dynamics of binaries in star clusters is a crucial ingredient to shed light on the formation channels of gravitational wave events \citep{abbott2016b}.

%%%% EXAMPLE OF A FIGURE %%%%%
%  \begin{figure}
%    \includegraphics[scale=0.55]{cantor1.eps}
%    \caption[Shortened figure caption for the list of illustrations]
%      {A Cantor repeller; figure captions are justified.}
%    \label{Author1Fig:cantor}
%     \end{figure}

%%%% An example of a Table 
%  \begin{table}[h!]
%    \begin{minipage}{180pt}
%    \caption[Shortened table caption for the list of tables]
%      {Longer table captions have to be placed inside
%      a minipage, otherwise they overhang the table rules.}
%    \label{Author1Table:example}
%    \addtolength\tabcolsep{2pt}% to stretch columns, if required
%      \begin{tabular}{@{}c@{\hspace{25pt}}ccc@{}}
%        \hline \hline
%        Figure\footnote{\textit{Note:} You must also use a minipage
%          environment if you have footnotes.}
%          & $hA$ & $hB$ & $hC$\\
%        \hline
%        1 & $\exp\left(\pi i\frac58\right)$
%          & $\exp\left(\pi i\frac18\right)$ & $0$\\[3pt]
%        2 & $-1$    & $\exp\left(\pi i\frac34\right)$ & $1$\\[12pt]
%        3 & $-4+3i$ & $-4+3i$ & $\frac74$\\[3pt]
%        4 & $-2$    & $-2$    & $\frac54 i$ \\
%        \hline \hline
%      \end{tabular}
%    \end{minipage}
%  \end{table}

  \bibliography{percolation}\label{refs}

\begin{thebibliography}{8}
  \expandafter\ifx\csname natexlab\endcsname\relax
  \def\natexlab#1{#1}\fi
  \expandafter\ifx\csname selectlanguage\endcsname\relax
  \def\selectlanguage#1{\relax}\fi


\bibitem[Abbott et al.(2016a)]{abbott2016a} Abbott, B.~P., Abbott, R., Abbott, T.~D., et al.\ 2016, Observation of Gravitational Waves from a Binary Black Hole Merger, {\emph{Physical Review Letters}}, 116, 061102 
\bibitem[Abbott et al.(2016b)]{abbott2016b} Abbott, B.~P., Abbott, R., Abbott, T.~D., et al.\ 2016, Binary Black Hole Mergers in the First Advanced LIGO Observing Run, {\emph{Physical Review X}}, 6, 041015 
\bibitem[Abbott et al.(2017)]{abbott2017IMBH} Abbott, B.~P., Abbott, R., Abbott, T.~D., et al.\ 2017, Search for intermediate mass black hole binaries in the first observing run of Advanced LIGO, \prd, 96, 022001 
\bibitem[Abbott et al.(2017a)]{abbott2017a} Abbott, B.~P., Abbott, R., Abbott, T.~D., et al.\ 2017, GW170104: Observation of a 50-Solar-Mass Binary Black Hole Coalescence at Redshift 0.2, {\emph{Physical Review Letters}}, 118, 221101
\bibitem[Abbott et al.(2017b)]{abbott2017b} Abbott, B.~P., Abbott, R., Abbott, T.~D., et al.\ 2017, GW170814: A Three-Detector Observation of Gravitational Waves from a Binary Black Hole Coalescence, {\emph{Physical Review Letters}}, 119, 141101 
\bibitem[Abbott et al.(2017c)]{abbott2017c} The LIGO Scientific Collaboration, the Virgo Collaboration, Abbott, B.~P., et al.\ 2017, GW170608: Observation of a 19 Solar-mass Binary Black Hole Coalescence, \apjl{}, arXiv:1711.05578 
\bibitem[Allison et al.(2009)]{allison2009} Allison, R.~J., Goodwin, S.~P., Parker, R.~J., et al.\ 2009, Dynamical Mass Segregation on a Very Short Timescale, \apjl, 700, L99 
\bibitem[Angeletti \& Giannone(1977)]{angeletti1977} Angeletti, L., \& Giannone, P.\ 1977, Dynamical evolution of clusters with stellar mass loss, \aap, 58, 363 
\bibitem[Angeletti \& Giannone(1980)]{angeletti1980} Angeletti, L., \& Giannone, P.\ 1980, \aap, Dynamical evolution of cluster models with a continuous stellar mass loss, 85, 113 
\bibitem[Askar et al.(2017)]{askar2017} Askar, A., Szkudlarek, M., Gondek-Rosi{\'n}ska, D., Giersz, M., \& Bulik, T.\ 2017, MOCCA-SURVEY Database - I. Coalescing binary black holes originating from globular clusters, \mnras, 464, L36 
\bibitem[Banerjee(2017)]{banerjee2017} Banerjee, S.\ 2017, Stellar-mass black holes in young massive and open stellar clusters and their role in gravitational-wave generation, \mnras, 467, 524 
\bibitem[Beccari et al.(2006)]{beccari2006} Beccari, G., Ferraro, F.~R., Possenti, A., et al.\ 2006, The Dynamical State and Blue Straggler Population of the Globular Cluster NGC 6266 (M62), \aj, 131, 2551 
\bibitem[Belczynski et al.(2010)]{belczynski2010} Belczynski, K., Bulik, T., Fryer, C.~L., et al.\ 2010, On the Maximum Mass of Stellar Black Holes, \apj, 714, 1217
\bibitem[Belczynski et al.(2016)]{belczynski2016} Belczynski, K., Holz, D.~E., Bulik, T., \& O'Shaughnessy, R.\ 2016, \nat, 534, 512 
\bibitem[Bianchini et al.(2016)]{bianchini2016} Bianchini, P., van de Ven, G., Norris, M.~A., Schinnerer, E., \& Varri, A.~L.\ 2016, A novel look at energy equipartition in globular clusters, \mnras, 458, 3644 
\bibitem[Boltzmann(1876)]{boltzmann1876} Boltzmann, L (1876). \"Uber die Natur der Gasmolek\"ule (On the nature of gas molecules). Wiener Berichte (in German). 74: 553–560.
\bibitem[Bonnell \& Davies(1998)]{bonnelldavies1998} Bonnell, I.~A., \& Davies, M.~B.\ 1998, Mass segregation in young stellar clusters, \mnras, 295, 691 
\bibitem[Chernoff \& Weinberg(1990)]{chernoff1990} Chernoff, D.~F., \& Weinberg, M.~D.\ 1990, Evolution of globular clusters in the Galaxy, \apj, 351, 121  
\bibitem[Colpi et al.(2003)]{colpi2003} Colpi, M., Mapelli, M., \& Possenti, A.\ 2003, Probing the Presence of a Single or Binary Black Hole in the Globular Cluster NGC 6752 with Pulsar Dynamics, \apj, 599, 1260
\bibitem[Davies(2002)]{davies2002} Davies, M.~B.\ 2002, Omega Centauri, A Unique Window into Astrophysics, 265, 215 
\bibitem[Downing(2012)]{downing2012} Downing, J.~M.~B.\ 2012, \mnras, 425, 2234 
\bibitem[Elson et al.(1987)]{elson1987} Elson, R., Hut, P., \& Inagaki, S.\ 1987, Dynamical evolution of globular clusters, \araa, 25, 565  
\bibitem[Farrell et al.(2009)]{farrell2009} Farrell, S.~A., Webb, N.~A., Barret, D., Godet, O., \& Rodrigues, J.~M.\ 2009, An intermediate-mass black hole of over 500 solar masses in the galaxy ESO243-49, \nat, 460, 73 
\bibitem[Ferraro et al.(2012)]{ferraro2012} Ferraro, F.~R., Lanzoni, B., Dalessandro, E., et al.\ 2012, Dynamical age differences among coeval star clusters as revealed by blue stragglers, \nat, 492, 393 
\bibitem[Fujii \& Portegies Zwart(2014)]{fujii2014} Fujii, M.~S., \& Portegies Zwart, S.\ 2014, The moment of core collapse in star clusters with a mass function, \mnras, 439, 1003 
\bibitem[Gebhardt et al.(2005)]{gebhardt2005} Gebhardt, K., Rich, R.~M., \& Ho, L.~C.\ 2005, An Intermediate-Mass Black Hole in the Globular Cluster G1, \apj, 634, 1093 
\bibitem[Giacobbo et al.(2018)]{giacobbo2018}Giacobbo, N., Mapelli, M., \& Spera, M.\ 2018, Merging black hole binaries: the effects of progenitor's metallicity, mass-loss rate and Eddington factor, \mnras, 474, 2959
\bibitem[Giersz et al.(2015)]{giersz2015} Giersz, M., Leigh, N., Hypki, A., L{\"u}tzgendorf, N., \& Askar, A.\ 2015, MOCCA code for star cluster simulations - IV. A new scenario for intermediate mass black hole formation in globular clusters, \mnras, 454, 3150 
\bibitem[Heggie(1975)]{heggie1975} Heggie, D.~C.\ 1975, Binary evolution in stellar dynamics, \mnras, 173, 729 
\bibitem[H{\'e}non(1961)]{henon1961} H{\'e}non, M.\ 1961, Sur l'\'evolution dynamique des amas globulaires, {\emph{Annales d'Astrophysique}}, 24, 369 
\bibitem[Hills \& Fullerton(1980)]{hills1980} Hills, J.~G., \& Fullerton, L.~W.\ 1980, Computer simulations of close encounters between single stars and hard binaries, \aj, 85, 1281 
\bibitem[Hills(1983)]{hills1983} Hills, J.~G.\ 1983, The effect of low-velocity, low-mass intruders (collisionless gas) on the dynamical evolution of a binary system, \aj, 88, 1269
\bibitem[Hypki \& Giersz(2017)]{hypki2017} Hypki, A., \& Giersz, M.\ 2017, mocca code for star cluster simulations - VI. Bimodal spatial distribution of blue stragglers, \mnras, 471, 2537  
\bibitem[King(1966)]{king1966} King, I.~R.\ 1966, The structure of star clusters. III. Some simple dynamical models, \aj, 71, 64 
\bibitem[K{\i}z{\i}ltan et al.(2017)]{kiziltan2017} K{\i}z{\i}ltan, B., Baumgardt, H., \& Loeb, A.\ 2017, An intermediate-mass black hole in the centre of the globular cluster 47 Tuc, \nat, 542, 203 
\bibitem[Mapelli et al.(2004)]{mapelli2004} Mapelli, M., Sigurdsson, S., Colpi, M., et al.\ 2004, The Contribution of Primordial Binaries to the Blue Straggler Population in 47 Tuc, \apjl, 605, L29 
\bibitem[Mapelli et al.(2006)]{mapelli2006} Mapelli, M., Sigurdsson, S., Ferraro, F.~R., et al.\ 2006, The radial distribution of blue straggler stars and the nature of their progenitors, \mnras, 373, 361 
\bibitem[Mapelli et al.(2009)]{mapelli2009} Mapelli, M., Colpi, M., \& Zampieri, L.\ 2009, Low metallicity and ultra-luminous X-ray sources in the Cartwheel galaxy, \mnras, 395, L71 
\bibitem[Mapelli \& Bressan(2013)]{mapellibressan2013} Mapelli, M., \& Bressan, A.\ 2013, Impact of metallicity on the evolution of young star clusters, \mnras, 430, 3120 
\bibitem[Mapelli et al.(2013)]{mapelli2013} Mapelli, M., Zampieri, L., Ripamonti, E., \& Bressan, A.\ 2013, Dynamics of stellar black holes in young star clusters with different metallicities - I. Implications for X-ray binaries, \mnras, 429, 2298 
\bibitem[Mapelli(2016)]{mapelli2016} Mapelli, M.\ 2016, Massive black hole binaries from runaway collisions: the impact of metallicity, \mnras, 459, 3432
\bibitem[Mapelli et al.(2017)]{mapelli2017} Mapelli, M., Giacobbo, N., Ripamonti, E., \& Spera, M.\ 2017, The cosmic merger rate of stellar black hole binaries from the Illustris simulation, \mnras, 472, 2422  
\bibitem[Maxwell(1872)]{maxwell1872}Maxwell, J. C., Theory of Heat (New York: D. Appleton \&  Co., 1872)
\bibitem[Merritt(1981)]{merritt1981} Merritt, D.\ 1981, Two-component stellar systems in thermal and dynamical equilibrium, \aj, 86, 318 
\bibitem[Miller \& Hamilton(2002)]{miller2002} Miller, M.~C., \& Hamilton, D.~P.\ 2002, Production of intermediate-mass black holes in globular clusters, \mnras, 330, 232 
\bibitem[Miocchi(2006)]{miocchi2006} Miocchi, P.\ 2006, Central energy equipartition in multimass models of globular clusters, \mnras, 366, 227 
\bibitem[Parker et al.(2016)]{parker2016} Parker, R.~J., Goodwin, S.~P., Wright, N.~J., Meyer, M.~R., \& Quanz, S.~P.\ 2016, \mnras, 459, L119 
\bibitem[Peters(1964)]{peters1964} Peters, P.~C.\ 1964, Gravitational Radiation and the Motion of Two Point Masses, {\emph{Physical Review}}, 136, 1224 
\bibitem[Portegies Zwart \& McMillan(2002)]{portegieszwart2002} Portegies Zwart, S.~F., \& McMillan, S.~L.~W.\ 2002, The Runaway Growth of Intermediate-Mass Black Holes in Dense Star Clusters, \apj, 576, 899 
\bibitem[Portegies Zwart et al.(2004)]{portegieszwart2004} Portegies Zwart, S.~F., Baumgardt, H., Hut, P., Makino, J., \& McMillan, S.~L.~W.\ 2004, \nat, 428, 724 
\bibitem[Portegies Zwart et al.(2010)]{portegieszwart2010} Portegies Zwart, S.~F., McMillan, S.~L.~W., \& Gieles, M.\ 2010, Young Massive Star Clusters, \araa, 48, 431
\bibitem[Quinlan(1996)]{quinlan1996} Quinlan, G.~D. 1996, \na, 1, 35 
\bibitem[Rodriguez et al.(2016)]{rodriguez2016} Rodriguez, C.~L., Chatterjee, S., \& Rasio, F.~A.\ 2016, Binary black hole mergers from globular clusters: Masses, merger rates, and the impact of stellar evolution, \prd, 93, 084029 
\bibitem[Schulman et al.(2012)]{schulman2012} Schulman, R.~D., Glebbeek, E., \& Sills, A.\ 2012, \mnras, 420, 651 
\bibitem[Servillat et al.(2011)]{servillat2011} Servillat, M., Farrell, S.~A., Lin, D., et al.\ 2011, \apj, 743, 6 
\bibitem[Spera et al.(2015)]{spera2015} Spera, M., Mapelli, M., \& Bressan, A.\ 2015, The mass spectrum of compact remnants from the PARSEC stellar evolution tracks, \mnras, 451, 4086 
\bibitem[Spera et al.(2016)]{spera2016} Spera, M., Mapelli, M., \& Jeffries, R.~D.\ 2016, Do open star clusters evolve towards energy equipartition?, \mnras, 460, 317 
\bibitem[Spera \& Mapelli(2017)]{spera2017} Spera, M., \& Mapelli, M.\ 2017, Very massive stars, pair-instability supernovae and intermediate-mass black holes with the sevn code, \mnras, 470, 4739 
\bibitem[Spitzer(1969)]{spitzer1969} Spitzer, L., Jr.\ 1969, \apjl, 158, L139 
\bibitem[Spitzer \& Hart(1971)]{spitzer1971} Spitzer, L., Jr., \& Hart, M.~H.\ 1971, \apj, 164, 399 
\bibitem[Spitzer(1987)]{spitzer1987} Spitzer, L.\ 1987, Princeton, NJ, Princeton University Press, 1987, 191 p.,  
\bibitem[Trani et al.(2014)]{trani2014} Trani, A.~A., Mapelli, M., \& Bressan, A.\ 2014, \mnras, 445, 1967 
\bibitem[Trenti \& van der Marel(2013)]{trenti2013} Trenti, M., \& van der Marel, R.\ 2013, \mnras, 435, 3272 
\bibitem[Vink et al.(2001)]{vink2001} Vink, J.~S., de Koter, A., \& Lamers, H.~J.~G.~L.~M.\ 2001, Mass-loss predictions for O and B stars as a function of metallicity, \aap, 369, 574 
\bibitem[Vishniac(1978)]{vishniac1978} Vishniac, E.~T.\ 1978, \apj, 223, 986 
\bibitem[Ziosi et al.(2014)]{ziosi2014} Ziosi, B.~M., Mapelli, M., Branchesi, M., \& Tormen, G.\ 2014, Dynamics of stellar black holes in young star clusters with different metallicities - II. Black hole-black hole binaries, \mnras, 441, 3703 




%\bibitem[Heggie, 1975]{Heggie1975}
%  Heggie, D. C. 1975.
%  Binary evolution in stellar dynamics.
%  {\em MNRAS}, {\bf 173}, 729--787.
% \end{thebibliography}
\end{thebibliography}
  \bibliographystyle{cambridgeauthordate}
  
 %%%%%%%%%%%%%%%%%%%%%%%%%%% 
 \copyrightline{} 
 \printindex
    %this is to check if you are happy with your index -- it will not appear here in the book and you can ignore the additional page
    
%%%%%%%%%% END %%%%%%
\end{document}